# Tangled String for Multi-Scale Explanation of Contextual Shifts in Stock Market

**Yukio Ohsawa[1], Teruaki Hayashi[1], and Takaaki Yoshino [2]**

[1]  Affiliation 1; [2]    School of Eng., The University of Tokyo, Affiliation 2; Nissei Asset Management Corp.
*  Correspondence: info@panda.sys.t.u-tokyo.ac.jp; 7-3-1 Hongo, Bunkyo-ku Tokyo 113-8656

**Abstract:** The original research question here is given by marketers in general, i.e., how to explain the changes in the desired timescale of the market. Tangled String, a sequence visualization tool based on the metaphor where contexts in a sequence are compared to tangled pills in a string, is here extended and diverted to detecting stocks that trigger changes in the market and to explaining the scenario of contextual shifts in the market. Here, the sequential data on the stocks of top 10 weekly increase rates in the First Section of the Tokyo Stock Exchange for 12 years are visualized by Tangled String. The changing in the prices of stocks is a mixture of various timescales and can be explained in the time-scale set as desired by using TS. Also, it is found that the change points found by TS coincided by high precision with the real changes in each stock price. As TS has been created from the data-driven innovation platform called Innovators Marketplace on Data Jackets and is extended to satisfy data users, this paper is as evidence of the contribution of the market of data to data-driven innovations.

## 1. Introduction: The problem definition

Dealers and analysists of stocks are often classified simply into fundamentalists and technical analysts i.e., chartists. The former consider the basic effects of events significantly affecting stock dealers' decisions, whereas the latter analyze the chart i.e. the time series of stock prices. Although these two tend to be regarded as different types of dealers, it is quite natural that a chartist considers the effects of social events in data other than charts [1] and that a fundamentalist may refer to the sequential data of stock prices [2, 3],. This existence of two approaches is partially similar to the positioning of real (human) analysts' and machine learning. Tools of machine learning are recently adopted to forecast demands in markets [4,5,6] from data, regarded as partial automation of technical analysts' thoughts.  However, the changes in the market due to external events are hard to explain by just learning patterns from data, because external events mean fundamental causes of dealers' decisions that may not be included in available data.  Not only dealers but also political planners of the market need not only to forecast but also to explain the changes, i.e., to relate a change in the observation with its causes that may be external events.

Suppose we have stock market data $D$ give as $\{B_t \mid 0 < t \leq T\}$, where $B_t$ stands for a *basket* (i.e., a set of members of itemset $I$, that are items or words forming a unit of co-occurrence such as a shopping basket in a supermarket, a sentence in a document, or a set of top-rank up-pricing stocks) in the high rank preference of dealers at time $t$ and $T$ is the sequence length of the data. Such basket data has been used for explaining patterns and changes in the market [7,8,9].  For example, suppose the weekly average prices of stocks of firms in healthcare businesses are increasing the most in the stock market for a month, whereas those on electronics were more overwhelming last year.  The cause of this new trend may be the interest of dealers in healthcare due to a movie shown at the end of last year appealing the negative effect of daily foods on human's health. This causality may be explained if the stock analyst, who broadcasts his analysis to dealers, have been paying attention to movies (that is certainly not in set $I$) during in the corresponding period. Then, the analyst may propose a strategy to promote the stocks of healthcare to stock dealers for the period the movie is on the show. Furthermore, the political section of the nation can plan a policy to support healthcare businesses if the new trend is regarded as sustainable in a long timescale.

In the approach of machine learning, change points have been detected on the changes in the features (parameters in parametric models) of models of time series. By projecting data to principal components, for example, the sensitivity of change detection is sharpened if the method is used for detecting the changes in the correlation, the variance, and the mean, of components with time [10]. Changes in parameters in the model capturing the structure of latent causality have been detected in discrete [11, 12, 13] and continuous [14] time series and the method for the latter has been shown to work also for the former. Here suppose the changes in the values of the parameter set $\Theta$, from time $t-\delta t$ to $t$, is learned as $\Theta[t-\Delta t, t] - \Theta[t-\delta t-\Delta t, t-\delta t]$ where $\Delta t$ is the length of the training time of the data to learn $\Theta$ from, and $\delta t$ is the given time step from before ($t-\delta t$) to present ($t$) to be compared. The precision of change detection is expected to be the better for the larger $\Delta t$ that can be regarded as a part of tolerant delay i.e., the length of time the analysis should wait for detecting a change. In [8], the author pointed out that needing a large $\Delta t$ is not convenient from the viewpoint to explain the change quickly. On the other hand, in this paper, the focus is to detect events in a sequence to provide a meaningful explanation of the change fitting the timescale of an analyst's or a dealer's purpose. For the machine learning viewpoint above, this corresponds to regarding the change points detected for a value of $\delta t$ given for the users' purpose, rather than $\Delta t$, as useful information of which even the lack of accuracy may be allowed by adding external information to explain the change.

Not only to detect but also to explain a change, the dealer may consider employing methods to learn latent topics from sequential data. Consecutive time segments, each of which is relevant to a vector in the space of a limited number of latent topics (without known labels corresponding to the topics) are obtained by the dynamic topic model (DTM) [15]. By applying DTM to sales data, the changes in consumers' interests can be detected as the boundaries between the obtained time segments corresponding to discontinuous changes in the topic vector. Topic tracking model (TTM) has been also presented to analyze the change of each consumer if the behavior of consumer $c$ is reflected on data in the form of $B_{t,c}$ instead of $B_t$ above [9]. The hierarchical structure of topics has also been learned [16]. These topic models have a potential to explain changes behind observed events with linking to underlying topics that are a kind of external information representing the trends for periods in the time series. On the other hand, the aim of this paper is to cope with a target sequence which may be a mixture of multiple event sequences of various timescales, that may be explained differently as various scenarios represented by transitions from/to trends triggered by events at changing points. For example, the sequence of stock prices for a decade can be explained as the shifts of trendy business domains if we look by the yearly timescale, but we may find different change points if we look in the timescale of months such as the effects of politicians' broadcasted words on dealers' attitudes that differ from the industrial trend shifts. Thus, we can now regard $D$ given as $\{B_{i,t}|\ i \in [1,n], t \in [1,T]\}$ in place of $\{B_t|\ 0 < t \le T\}$ mentioned above where $i$ stands for the timescale and $B_t$ is the union of $B_{i,t}$ for all $i$ in {1, 2, …, n}. Note there are intersections among $B_{i,t}$ for different $i$. Thus the problem here is to obtain $\{ent_{i,t}|\ i \in [1,n], t \in [1,T]\}$ and $\{ext_{i,t}|\ i \in [1,n], t \in [1,T]\}$ where $ent_{i,t}$ and $ext_{i,t}$ respectively represent events at time $t$ that worked as an entrance to and an exit from a trend for the $i$-th timescale considered. If no event at time $t$ played the role, $ent_{i,t}$ and $ext_{i,t}$ both should take NULL as the value.

However, if a tool just analyzes to show these sets as outputs, the links between them can not be shown, so the explanation is still hard for human analyst or fundamentalists. Methods for data visualization may become a key technology for integrating the approaches of fundamentalist and technical analysts, by providing both classes of (if we can classify) dealers with desired information to refer to. For example, a sheer chart of stock price may turn out to be a useful visualization tool in this sense, if essential tipping points of price transition are shown so that a dealer can combine the information with external information about events relevant to the stock market. As well, the graphical visualization of the map of stocks and words in news provide hints about positive/negative moods of people in the market to events relevant to stocks [17,18].

In this paper, we extend and apply the Tangled String (TS), that deals with a sequence where patterns are disordered, i.e., patterns of various timescales co-exist that makes trends and the boundaries of trends hard to identify. Such a sequence is usual in a communication log of SNS like Twitter or in a market where the participants do not share the understanding of or are not explicitly aware of topics in the society or trends in the market. Behind the sequence of stock prices, various investors really have different reasons for buying and/or selling stocks reflecting their various ways to understand the market on their interest in the social dynamics in various time-scales. In such a case, a segment, if formed, cannot be featured by a certain typical pattern or a certain distribution of word frequencies. By use of TS, not only the output sets mentioned above are obtained but also the connections of a trend to preceding and subsequent trends are visualized.

## 2. Tangled String Invented from The Market of data

In this section, we introduce Tangled String (TS hereafter [19, 20]) as a method for the analysis and visualization of sequential data. This tool is expected to fulfill the purpose stated in the previous section, i.e., to segment a sequence where various patterns co-exist, that appear repeatedly with changing the occurrence order of events.

### A. The origin of Tangled String

The original version of TS has been a by-product of Innovators Marketplace on Data Jackets (IMDJ), a market-like creativity support system [21, 22]. In IMDJ, data providers show data and/or possible methods for data analysis, and data users cast requirements for data and for analysis technologies. TS has been created based on providers' response to the requirements presented by potential data users in IMDJ. Events in given sequential data are analyzed and visualized by TS, where the importance of each event is computed according to on their positions in the sequence. This visualization has been designed to fit users' requirements presented in IMDJ. In order to realize the method $M_1$ below for satisfying the requirement $R_1$, TS has been invented for ones who desire to detect a message of the highest social impact from messages in SNS about a disaster ($DJ_1$) with considering the contexts in the preceding and subsequent messages. After choosing such words, additional information about disasters ($DJ_2$) will be used for validating the credibility of the message corresponding to the chosen words.

- **Requirement** $R_1$: collect information useful for business decision
- **Method** $M_1$: evaluate the impact of a message in a sequence and link it to external information
- **Data for realizing** $M_1$: {$DJ_1$: log text of communication, $DJ_2$: information about disasters}

The performance of TS has been shown to satisfy requirements such as of $R_1$ above and others in the analysis of the text in timelines and the sensor-based log of human's movements in office [19]. TS also came to be required as in the following set.

- **Requirement** $R_2$: detect events in the tipping points of consumer behaviors in the market
- **Method** $M_2$: obtain a high-impact event in a sequence by TS and combine it with external information to explain the causality
- **Data for realizing** $M_2$: {$DJ_3$: log of consumptions or purchase history, $DJ_4$: social events and news}

In this paper, we take the stock market as the target market in $R_2$ and stock dealers as consumers. If we regard this as a sheer application of TS, the novelty here is that we introduce the analogy from the communication log in $M_1$ to the consumption log in $M_2$, and from the requirement-method-data relation for $M_1$ to the one for $M_2$ for fitting the purpose in $R_2$.

In TS, a sequence is modeled by a string, of which each entangled part is called a *pill*. A pill is defined as a sequence of events, from an event (the appearance of an item) to the appearance of the same item in the sequence. A pill may include other pills, and a maximum pill that is not included by any other pill is called just a pill hereafter. In order to reduce ambiguity, let us call the symbol expressing an item (a stock that priced up in this paper) a *token*. In natural language text, the same token may mean different items, and the same item may be represented by different tokens. On the other hand, when we mean the entity meant by the token, the entity is called an *item*. In this paper, each item represents a stock of the ID number (a token representing a unique stock) and its increase in the price is called an event. Therefore, an item and a token correspond to each other uniquely, and an event is an appearance of an item at a certain time (or a token at a certain position) in a sequence.

Pills formed on the sequence may meet each other, due to sharing the same tokens. As a result, a string having pills may cross with itself to make a larger pill. For example, see the substring between $w_1'$ and $w_1$ Fig.1. Here the two pills including $\{w_1', s_2\}$ and $\{w_1, s_6\}$ respectively get jointed at $s_2$ and $s_6$ and form one pill because tokens for $s_2$ and $s_6$ are the same. Events in the sequence are classified as a result into two: (1) events in pills and (2) others on *wires* that are substrings connecting pills. A token may appear multiple times in a pill within a constant time window, that means an item repeatedly appeared in the trend corresponding to the pill. On the other hand, events on a wire represent the essential contextual flow, from a trend represented by a pill to the next trend. In this sense, the connections of pills via wires are expected to be used to explain changes in the sequence. By the structure made of pills and wires, we aim to visualize sequential data for enabling a user to interpret and explain the contextual or the causal flow. Below let us show the algorithm of TS, where the cycles for all the events in the sequence follow the initial setting step.

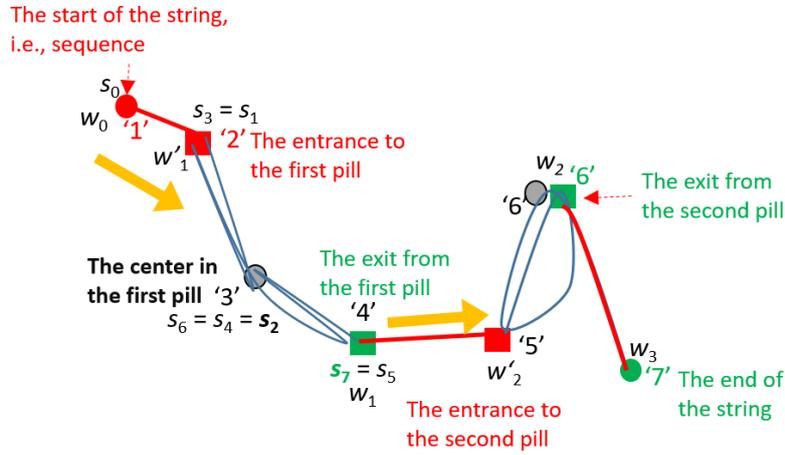

**Figure 1.** An example of Tangled String allied for *String*.

***The original version algorithm of Tangled String*** (Revised from [19] without changing the meaning)

Initial setting:

$String = \{s_1, \ldots, \ldots s_i, \ldots s_j, \ldots, \ldots, \quad s_L\}$,

$Wire = \{w_1, w_2, w_3, \ldots, w_L\} = \{s_1, s_2, s_3, \ldots, s_L\}$

for $i = 1$ to $L$:      $pill(s_i) = s_i$   where $pill(s)$ means the pill including event $s$.

                       $pill\_weight(s_i) = wire\_weight(s_i) = 0$

For $W$, a preset window, execute the cycles below:

For $i = W + 1$ to $L$:

     $Neighbors\ (s_i, W) = \{s_{i-W}, s_{i-W+1} \ldots, \ldots, s_{i-1}\}$

         if $\exists_{s_{j(<i)} \in Neighbors(s_i, W)} \{token(s_i) = token(s_j)\}$:

                 $j = min\{\ j\ |s_j \in Neighbors(s_i, W), token(s_i) = token(s_j)\}$

                 $r(s_i) = r(s_j)$                      #place $s_i$ at $r(s_j)$, the same position as $s_j$

                 $Wire = Wire\backslash\{s_j, s_{j+1}, \ldots, s_{i-1}\}$      # cut the subsequence from $s_j$ to $s_{i-1}$ from $Wire$

                 $pill(s_i) = \ldots = pill(s_{j+1}) = pill(s_j)$    # put all the events on the loop made by $s_i$ and $s_j$ in the same pill

                 $pill\_weight(s_j) += i - j$          # the length of the loop is added to the weight of event $s_j$ in the pill

         else    $r(s_i) = r(s_{i-1}) + a\ \{r(s_{i-1}) - r(s_{i-2})\}$    # place $s_i$ in the extension of the line from $s_{i-1}$ to $s_i$ ($a$: a real constant)

         end if

     end for

     for each *pill*

         $s_{ent}, s_{ext} =$ the first and the last event in *pill*

         $wire\_weight(s_{ent}) = wire\_weight(s_{ext}) = ext - ent$       # assign the pill size as the weight of each event on a wire

     end for

Here $s_i$ denotes an event that the symbol $token(s_i)$ appeared at the $j$-th position in the given string *String*, rather than the symbol itself. That is, $token(s_i)$ is unique to a symbol, and $s_i$ for multiple $i$'s may correspond to the same symbol. In the initial setting, no tangling but a single wire exists which is the given sequence. At this stage, therefore, events on the string and on the wire can be related by the pair-wise correspondence between each $w_k$ and $s_k$ for $k$ in $\{1, 2, \ldots, L\}$, where $L$ is the length of *String*.

In the cycles after the initial setting, each event $s_i$ in the string is taken one by one. For each $s_i$, event $s_j$ of the same token as $s_i$ is searched from the events preceding $s_i$ by within the time distance of $W$. Here, $W$ means the width of the window, given in weeks (e.g., $W = 3$ means the width is 3 weeks), represented by $Neighbors(s_i)$ in the backward search of events, i.e., the length of the time range [$i$ -$W$+1, $i$]. If such an event $s_j$ is found, a *pill* is obtained as the sub-string of *String* between $s_i$ in the procedure. $s_i$ and $s_j$ come to belong to the same pill (i.e., $pill(s_i) = pill(s_j)$). $s_i$ and $s_j$ are regarded as two appearances of the same token in the same trend. This trend is superficially observed as the appearance of the repeated set of tokens but does not necessarily mean the existence of a concrete reason for the appearance. Note a pill may finally become a part of a larger pill if some token in the succeeding subsequence (within $W$ baskets) matches with a token in the pill.

The substring between $s_j$ and $s_i$ is then pruned from *Wire*, that represents all the wires. As illustrated in Fig.1, the suffix number $k$ of $w_k$ may differ from $i$ of the corresponding node representing event $s_i$ because the wire between $s_j$ and $s_i$ comes to be deleted from *Wire* in the procedure, as illustrated in Fig.1. On the other hand, if no event such as $s_j$ of which the token ($token(s_i)$)) is equal to $token(s_i)$ found within the window of $W$ baskets before $s_i$, then $s_i$ and all the events in $Neighbors(s_i)$ will survive as a wire. After all the cycles, a position ($position_i$) in the 2D space is assigned to each $s_i$.

When a new event $s_i$ is added as a node in a pill $pill(s_j)$, the weight of the expanded pill denoted by $pill\_weight(s_j)$ is increased by $i - j$, that is the number of events added here to the pill. That is, the increase in the size of the pill due to absorbing $s_i$ is attributed to the contribution of the new path from $s_j$ to $s_j$ in the formation of the pill $pill(s_j)$. After all the cycles, the weights of $s_{ent}$ and $s_{ext}$ that are the first and the last event (i.e., $s_t$ of the smallest and the largest $t$) in each pill, denoted by $wire\_weight(s_{ent})$ and $wire\_weight(s_{ext})$ respectively, take the values of the size of the pill. This means the size of a pill is counted as the weight loaded on the edges of the wires connected to the pill. On these values of weight, the following key events are obtained:

*Key events in pills* (visualized as items in a pill): events $s$ of top $pill\_weight(s)$ values. This value tends to be the larger if $s$ is repeated the more frequently within $W$ consecutive events in a pill.

*Key events on wires* (visualized as enlarged red/green nodes at the entrances/exits of pills in the string): events ($u$'s) of top $wire\_weight(u)$ values. These events are not as frequent as key events in pills, but play an important role in explaining the contextual transition in the sequence, such as major trend shifts in the market. We call $s_{ent}$ the entrance to the pill, $s_{ext}$ the exit.

Thus, in TS, an event is regarded as a key event in a pill if the corresponding item is repeated in the pill, whereas an event which starts or halts a pill is regarded as a key event on a wire, or a wire-pill meeting point. For example, suppose TS is applied to the sequence *String* as Eq.(1).

$$String = 1\ ,2,\ 3,\ 2,\ 3,\ 4,\ 3,\ 4,\ 5,\ 6,\ 2,\ 5,\ 6,\ 7. \qquad (1)$$

In this case, as shown in Figure 1 and Figure 2(a), the event of the appearance of an item represented by token '2' starts the first pill including itself and '3' and '4', because '2' is the first token revisited by the string. That is, '2' is assigned to the event $s_1$ after the is also the third $s_3$. Token '3' is visited thrice i.e., as $\{s_2, s_4, s_6\}$ and '4' twice as $\{s_5, s_7\}$. In this pill, '4' is the last token repeated, regarded as the exit and is the start of the *wire* that is the connecting link to the next pill. Here, TS visualizes the flow of the sequence, highlighting the wires connecting a cluster (i.e. pill) to the next of frequent words. Events on the wires may not look outstanding because they are not "trendy" events in pills that appear frequently. However, the on-wire nodes play an essential role in the mainstream of the sequence because the connection in the whole structure is cut off if these events are lost. Although in this sense the event of token '3' in Figure 1 and Figure 2(a) should be a candidate to be a member of a wire because the structure becomes destroyed if we get rid of '3', here '3' is regarded not an on-wire but the main event in the first pill because the other events in the first pill are connected to '3.'

The window width $W$ plays an important role as a solution to the problem addressed in this paper because the setting $W$ to a larger value means to find the entrances to and the exits from pills of a longer timescale. Although the obtained time length of a pill is not in proportion to $W$, we can obtain an obvious change in the structure of the tangled string by changing $W$ and can find chancing points for different timescales. In the case of Figure 2 (a), '2' appears twice because the distance from the first and the second '2' is 7 in the sequence and the window width $W$ is set to 5. As in Figure 2 (b), on the other hand, the change in $W$ causes a radical change in the structure, if $W$ is set larger than the appearance period of the same item. Really, the structure changes from Figure 2 (a) to (b), losing the wire from '4' to '3' and unifying the two clusters in (a) into one in (b). However, the structure does not change more for a further increase in $W$.

An additional advantage of the algorithm is that the computational time is in the order of $O(L)$, i.e., linear to the length of the sequence. For the real use of TS, the position of each node and edge in the output of TS is revised by the wired stretch model [23], a specifically developed method for visualizing the graph with a starting and an ending node.

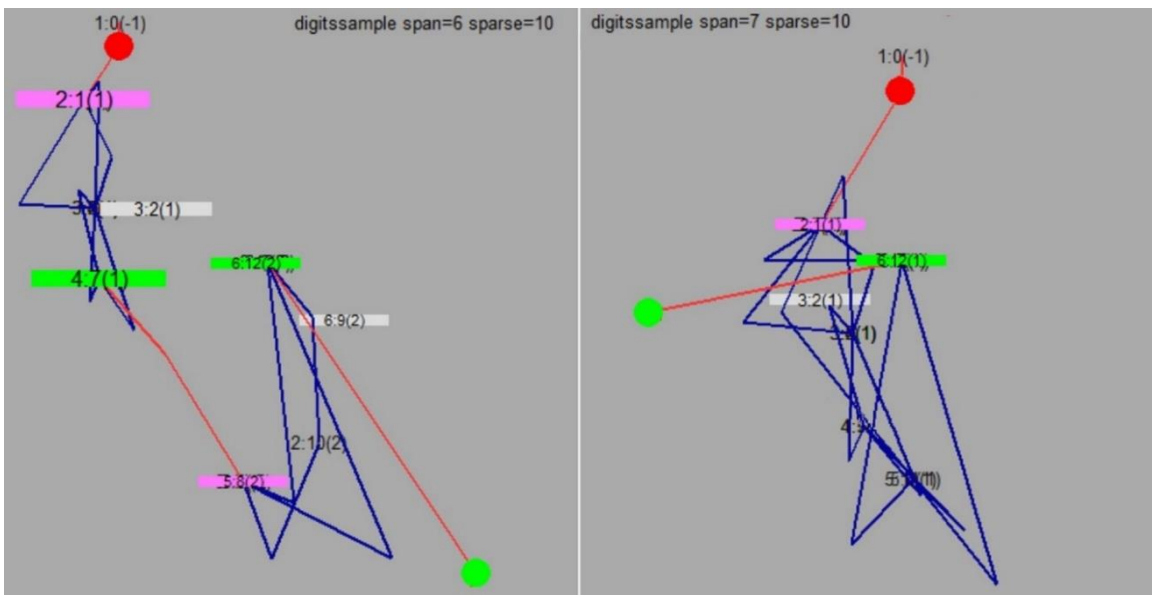

**Figure 2.** Tangled String employed for the widths ($W$) of 6 and 7 (the unit of $W$ is week).

### B. An extension of TS for dealing with basket data

*Tangled String for basket-set data*
Initial setting:
    *String* = {#$n_1$, $s_1$, $s_2$..., #$n_2$, $s..$, ...$s..$, ... , #$n_T$, , ... $s_L$},     where #n: the delimiter for each time *t*.
    *Wire* = {$w_1$, $w_2$, $w_3$, ..., $w_T$}               where $w_j$ is {#$n_j$, $s$, ..., $s_{mj}$ (just before #$n_{j+1}$)}
    for each *s* in *String*:        *pill*(s) = $w_j$       where $s_j$ is a member of $w_j$
                           $pill\_weight(s) = wire\_weight(s) = 0$
For $W$, a preset window, execute the cycles below:
    For each $s_i$ in *String*/{#$n_1$, #$n_2$, ... , #$n_T$ }:
        *Neighbors* ($s_i$, $W$) = ∪ $w_{k-W+1}$, ...,$w_k$ where $s_i$ is a member of $w_k$.
        if $^∃s_{j (<i)}$ ∈*Neighbors*($s_i$, $W$) s.t. *token*($s_i$) = *token*($s_j$):
                $j$= *min*{ $j$ |$s_j$ ∈*Neighbors*($s_i$, $W$), s.t. *token*($s_i$) = *token*($s_j$)}
                $r(s_i) = r(s_j)$                #place $s_i$ at $r(s_j)$, the same position as $s_j$
                $Wire = Wire$\{$w_p$, ...,$w_{k-1}$}     where $s_j$ is a member of $w_p$.
                $pill(s) = pill$ ($s_j$) for all $s$ in {$s_i$} ∪ all $s$ in {$w_p$, ...,$w_{k-1}$}
                $pill\_weight(s_j)$ += $i - j$      # the length of the loop is added to the weight of event $s_j$ in the pill
        else      $r(s_i) = r(s_{i-1}) + a$ {$r(s_{i-1})$ - $r(s_{i-2})$}   #place $s_i$ in the extension of the line from $s_{i-1}$ to $s_i$ ($a$: a real constant)
        end if
    end for
    for each *pill*
        $s_{ent}$, $s_{ext}$ = the first and the last event in *pill*
        $wire\_weight(s_{ent}) = wire\_weight(s_{ext}) = ext - ent$     # assign the pill size as the weight of each event on a wire
    end for

In the extension above, $s_i$ in the initial setting denotes the $i$-th of all events except "#n" in *String*, where *String* corresponds to the given target sequence and "#n" denotes delimiters between *baskets*. Other settings are similar to the original TS shown earlier. In this extended TS for sequential basket data, when the newest event $s_i$ is added, $s_j$ of the same token as $s_i$ is searched from the preceding neighbors of $s_i$ i.e., the nearest $W$ baskets before $s_i$. If such an event as $s_j$ is found, the substring between $s_j$ and $s_i$ becomes a pill as in Fig.1. Then, all the events in the same basket as $s_i$ join the same pill as all the events in the same basket as $s_j$. The entrance to the thus combined pill is $s_j$ if $s_j$ is the first event revisited in the pill, and the exit is the last revisited one ($s_i$, if $s_i$ is the last revisited) in the pill.

## 3. Results

The results presented here are two-fold. In 3.1, the outline of the data on weekly highly ranked price increasing stocks is shown. In 3.2, we show the coincidence of the entrances and exits of the pills obtained by TS with the beginning of the up-ward shift of the average price of stocks in the market for each $W$ from 4, 5, and 6. Also, the qualitative explanation of the trend shifts, given by an expert of stock analysis, is given for the output figures of Tangled String. Then in 3.3., we show the coincidence of the entrances and exits with the major changes in stock prices.

### 3.1. The data on up-pricing stocks

The Tangled String (TS hereafter) is here applied to the data on stock prices, of which the top 10 stocks of price increase rate per week are taken every week as (http://www.kabu-data.info/neagari/neagari_hizuke_w_1.htm). These data are taken for the First Section of the Tokyo Stock Market for 592 weeks from 6th July 2007 till 4th Jan 2019. The original data included the names of each company corresponding to a firm, but here we take just the stock ID number of each firm as in Table 1. Each set of 10 top stocks are taken as a basket as in Table 1. We also used data on Japan's stock-price trend (Nikkei Average) for evaluating the performance of TS in detecting the up/down-ward price changes in the stock market.

**Table 1.** The data for on weekly top up-pricing stock

|  |  | No. 1 | No. 2 | No. 3 | …, …, … | No.10 |
|---|---|---|---|---|---|---|
| 2007.7.6 | #n | 6378 | 8061 | 2678 | …, …, … | 2687 |
| 2007.7.13 | #n | 1907 | 6850 | 6316 | …, …, … | 1898 |
| … | #n | … | … | … | …, … | …, … |
| … | #n | 7600 | 4971 | 5302 | …, …, … | 8834 |
| … | #n | 1907 | 6378 | 7999 | …, …, … | 8934 |
| 2019.1.4 | #n | 4992 | 4344 | 6465 | …, …, … | 9501 |

### 3.2. The visualization of TS with changing the window width W

The visualizations by TS are shown from Figure 3 through Figure 5. Here, the letters in red and green respectively show $s_{ent}$ and $s_{ext}$ i.e., the events starting and ending pills. Figure (b) of each Figure 3, 4, and 5 include not all $s_{ent}$ and $s_{ext}$ but only relatively larger letters in (a) respectively i.e., of larger wire_weight shown in the algorithm, due to the constraint of the space. However, in both (a) and (b), the distance between a green node G and a red node R just after G turned out to be substantially smaller than between R and a green node G' just after R. That is, the size of a pill (from a red event to the next green one) is larger than a path of wire (from green to red) bridging a pill to the next pill.

The sequence visualized by TS in Figure 3 (a) setting window width $W$ to 4 corresponds to the time series of Nikkei Average in Figure 3 (b). Because 19 pills were obtained by TS here for the 11 y of the data, setting $W$ to 4 means to investigate pills by the time resolution of about 0.58 y. In this period, we do not find an obvious coincidence between the timing of red/green nodes and the long-term change points in the stock prices of Nikkei Average as far as we viewing the curve. On the other hand, the results in Figure 4 (a) and (b) are for setting $W$ to 5. Because we obtained 7 pills for the 11 y, $W = 5$ means the time resolution of about 1.6 y. In this case, we find a substantial coincidence between the timing of red/green nodes and the long-term changing in the Nikkei Average. That is, 5 of the 7 red nodes are within 3 months before the monotonic increase for 6 mo or longer in Nikkei Average in Figure 4 (b), 4 red nodes are within 1 mo before the increase of similar length, and 3 red nodes are within 1 mo before the beginning of a 1 y increase. Finally, the result in Figure 5 is for $W = 6$. Because we have 3 pills for the 11 y in (a), $W = 6$ means the time resolution of about 3.7 y. The two red nodes in Figure 5 (a) coincide with the months starting the monotonous increase in stock prices for 6 mo or longer, and one is the beginning of increase for 2 y. For the qualitative explanation of the changes considering the red/green nodes and more statistic evaluation of the coincidence of entrances to and exits from pills and relevant stocks, let us discuss in the next chapter.

### 3.3. The stock price changing and the two types of change points in TS

Next, we evaluated the coincidence of the entrances to and the exits from pills to the changing i.e., increase and decrease, of stock prices. For example, in Figure 6, the entrances and the exits tend to coincide with the starting and the halting of the lasting periods of the high-price trend. Figure 6 exemplifies a few of the cases where the prices of the stocks increased at the entrance. On the other hand, the price of the stock that started (appeared at the entrance of) a pill tended to decrease at the time of the exit from the started pill. To evaluate these tendencies, as in the left half of Table 2, let us compare the trend of time length $\Delta t$ before and after the entrances of pills, for stocks that appeared as entrances (red letters in TS: see Figure 7(g) for understanding). The comparison is also shown in the right half of Table 2 for stocks that appeared as exists (green letters in TS: see Figure 7(h)). Note we did not include cases where $W = 1$ or 2 in Table 2, because the number of pills came to be 132 and 93 respectively, meaning the average pill lengths were 1 mo and 1.4 mo i.e. shorter than any $\Delta t$ we chose (3 mo at least) for this evaluation. $W = 3$ is included here because 43 pills were obtained meaning the average pill length of 3.0 mo. In these results, the prices of both change-point stocks i.e., stocks appearing as entrances and as exits, increased substantially immediately after their appearance.

On the other hand, the probability by which the price of the stock at an entrance increased is relatively smaller at the time of the exit from a pill (as in Figure 7(b), (e), corresponding to (h)) than the price-up probability the of a stock at the time it appeared at the entrance (Figure 7(a), (d), corresponding to (g)). Also, for a stock that appeared at the entrance of a pill, the probability of the price-up after the time of the exit (as in Figure 7(c), (f), corresponding to (i)) is smaller than at the time of the entrance (Figure 7(a) and (d)). These tendencies stand especially for the larger value of $W$. This tendency is interpreted as in the discussions later.

In this paper we do not compare these results with methods for prediction with machine learning as in [4, 5, 6], because the aim to (1) detect the change points on which the future in/decrease trend can be forecasted in various time scales and also to (2) explain or predict the long-term trend from a change point in the stock market, by the same visualization method, has not been challenged by existing methods. Furthermore, the precision values obtained in Table 2 are higher than the prediction of up/down pricing of stocks by such machine learning techniques (e.g., 65% at the best in [4]) even though changes in the long-term trend precision is hard because of the uncertainty in the far future scenario. That is, in face of this difficulty, the precision of TS in the worst case to predict a middle/long-term (3 months or longer) increase in Table 2 (a) is 81% for the pill-entrance events, and 71% even for the increase larger than the standard deviation. These percentages are even higher than existing methods that achieved high accuracy in finding precursors to drawdowns of the market average [24], or the Rate of Change Oscillators as signs for buying/selling [25]. We also checked the tolerant delay in obtaining a changing point (entrance/exit), by entering data of time sequences of having length $dt$ after each entrance and each exit. As a result, for all the pill entrances and exits for 4 or larger $W$, we found the same entrances and exits where obtained for $dt$ set to 3 months or larger.

(a) Tangled string for window width *W* = 4 weeks

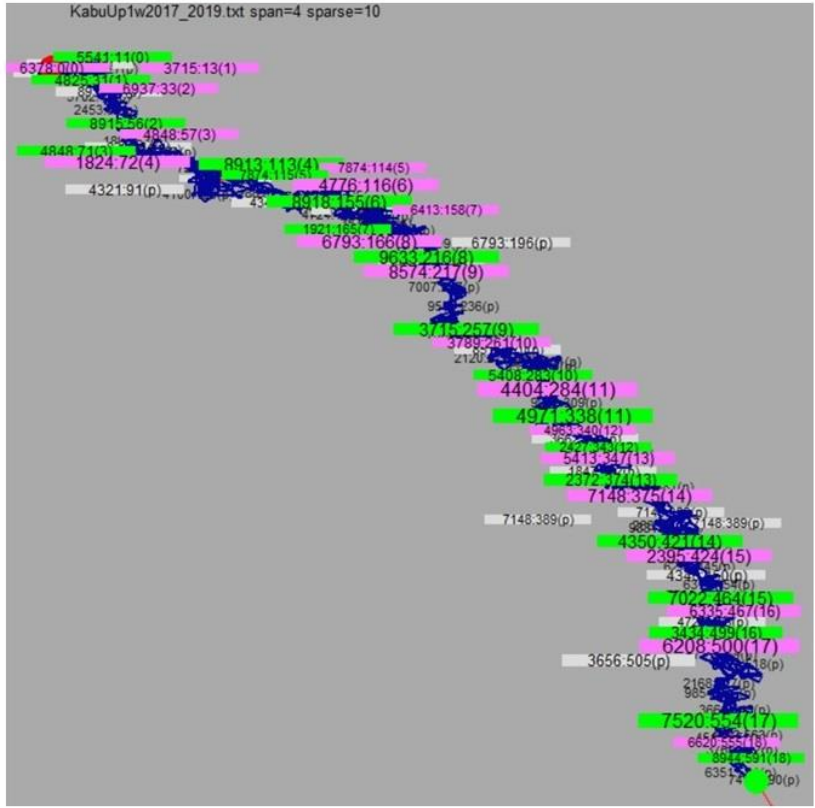

(b) The Nikkei Average chart with the stocks in the tangled string for *W* = 4 weeks

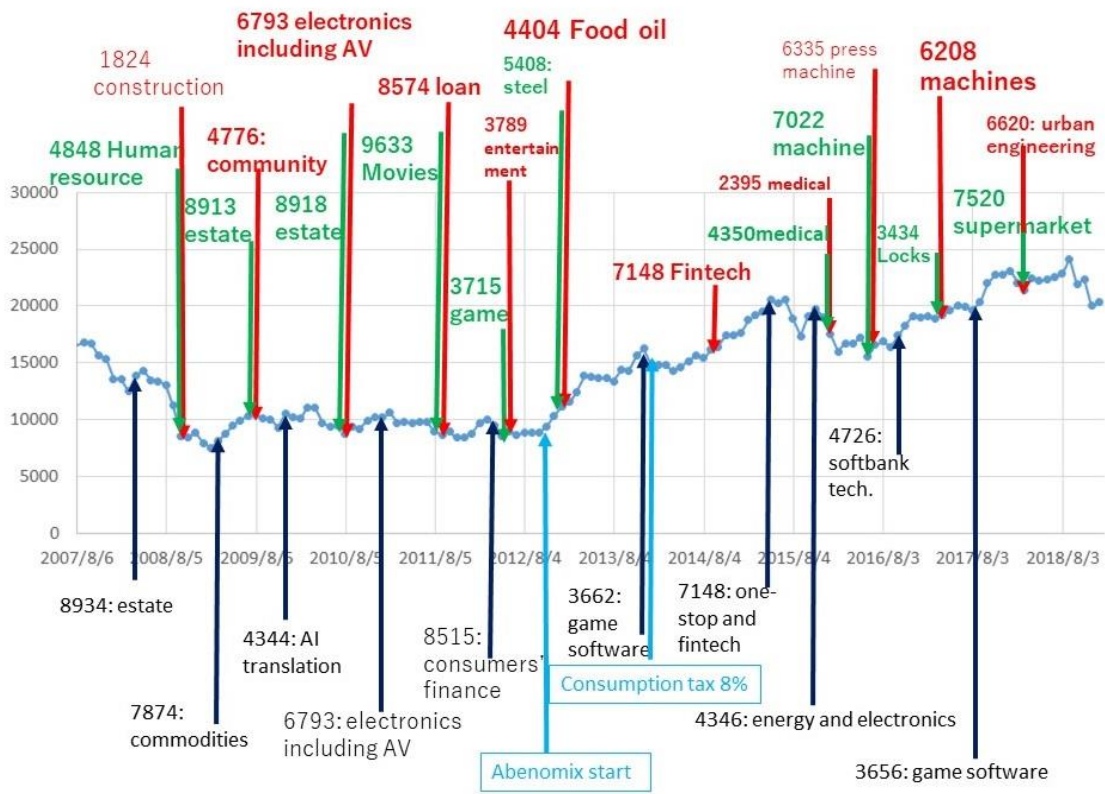

**Figure 3.**The result of TS for *W* = 4 (weeks). In (a), each red/green widget show X:Y:(Z) where X, Y, and Z represents the stock ID, its time (the week counted from the beginning of the target sequence) and the number of pills counted from the beginning respectively.

(a) Tangled string for *W* = 5 weeks

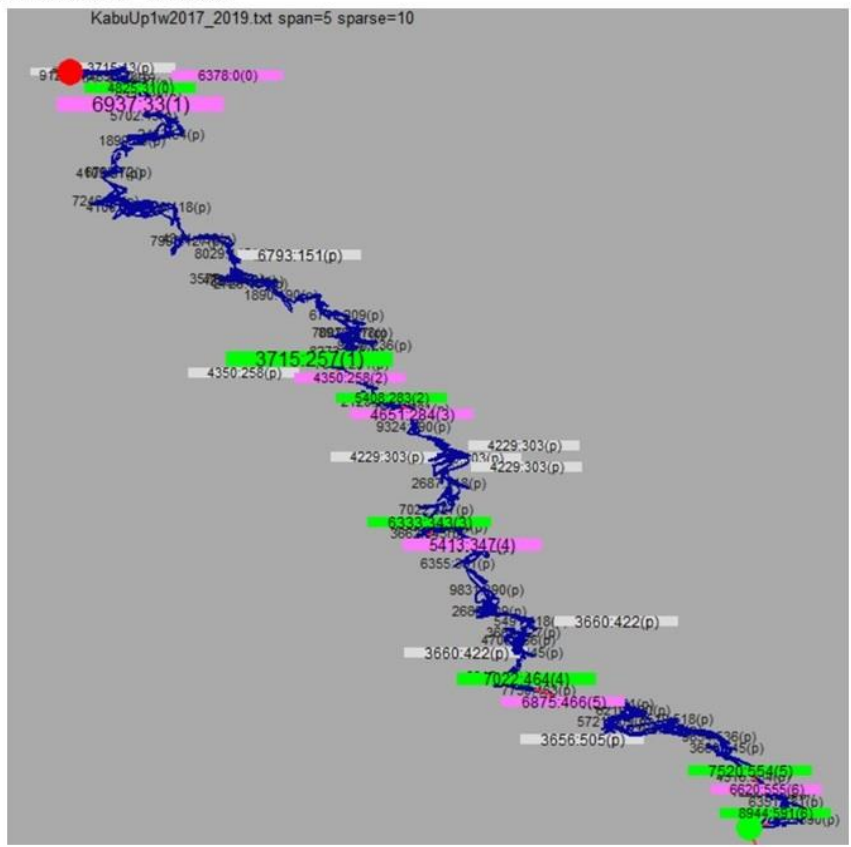

(b) The Nikkei Average chart with the stocks in the tangled string for *W* = 5 weeks

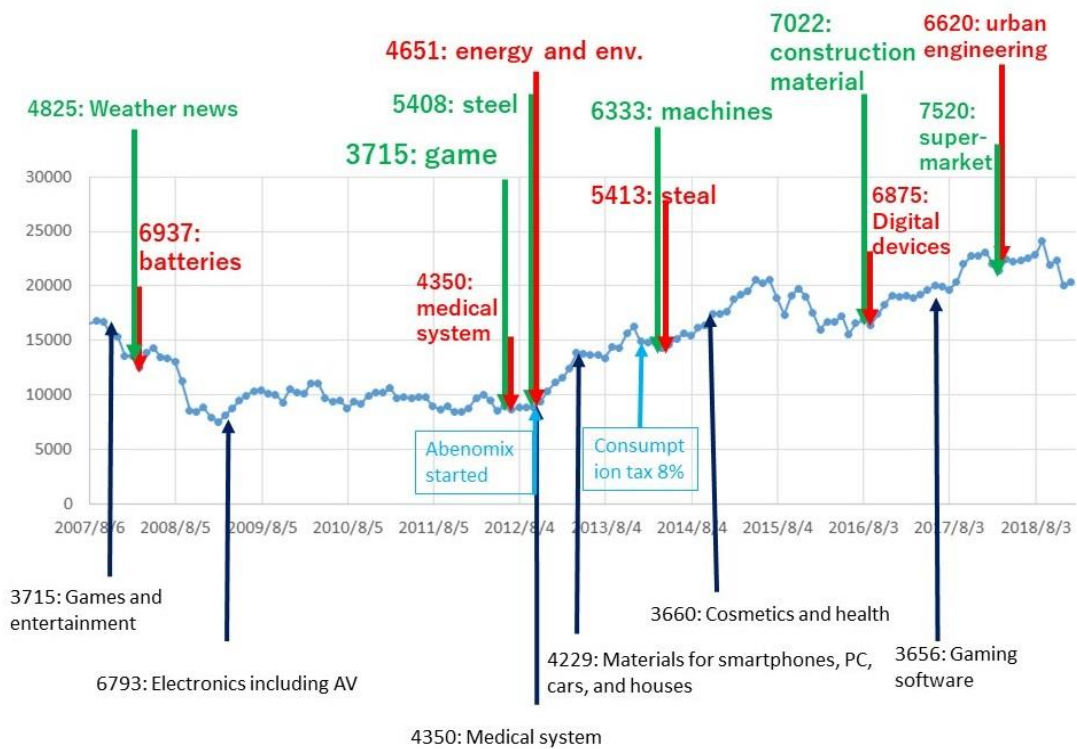

**Figure 4.** The result of TS for *W* = 5 (weeks).

(a) Tangled string for *W* = 6 weeks

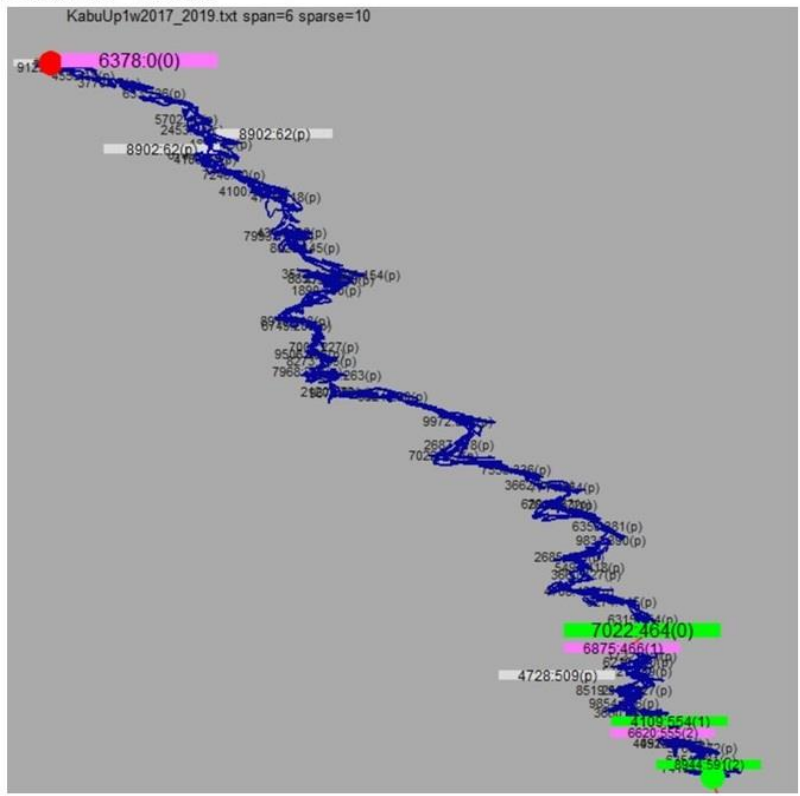

(b) The Nikkei Average chart with the stocks in the tangled string for *W* = 6 weeks

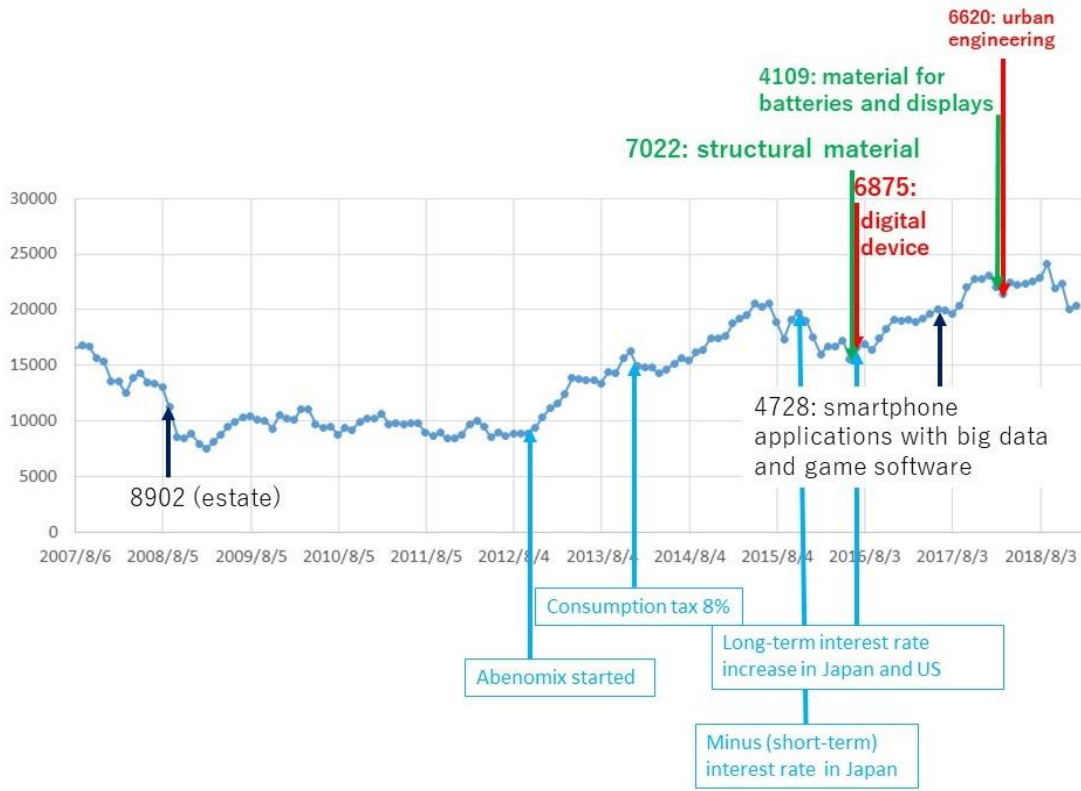

**Figure 5.** The result of TS for *W* = 6 (weeks).

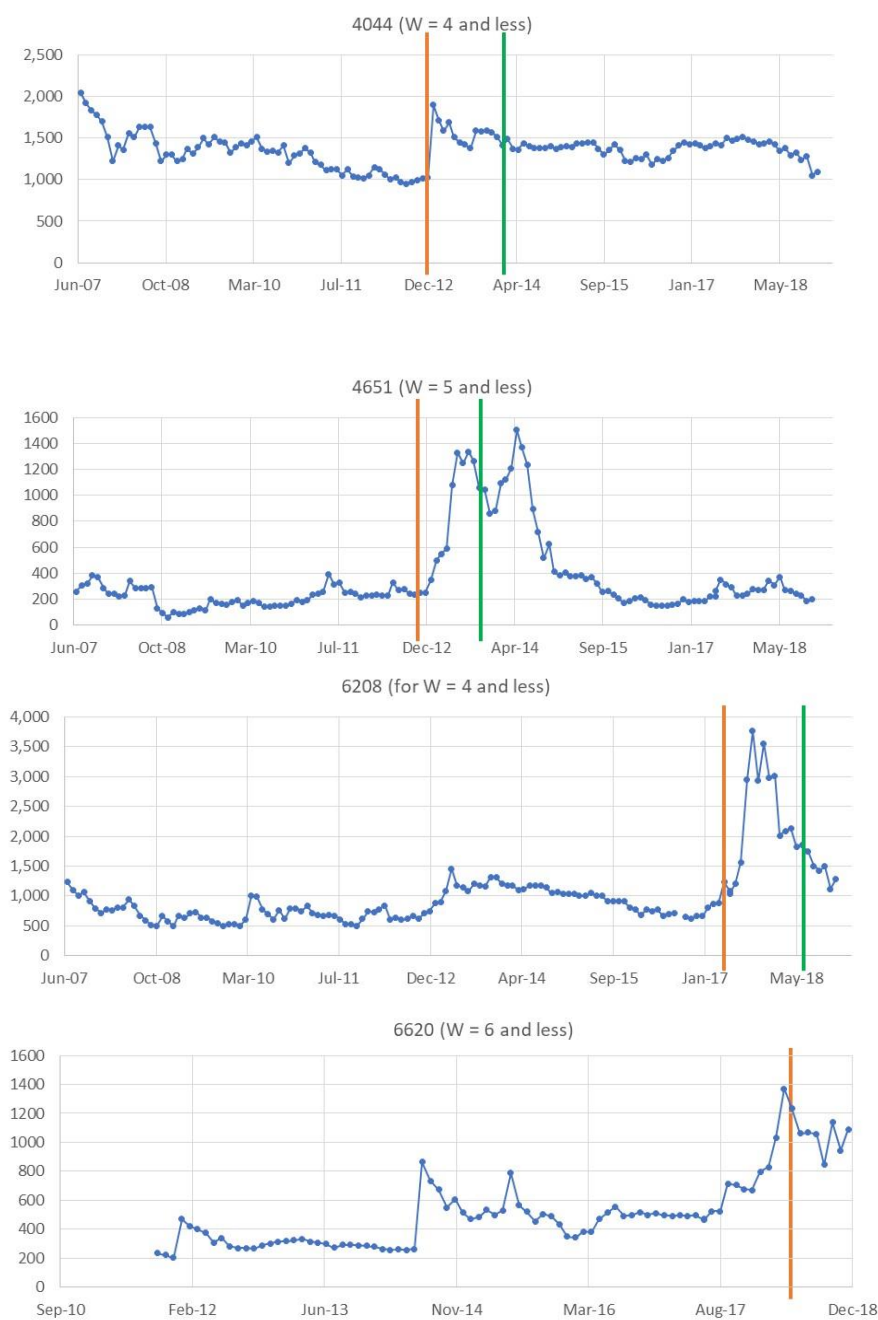

**Figure 6.** Examples of pill entrances and exits, compared with changes in stock prices

**Table 2**. The total counts of the increase/decrease of stock prices at the entrances or the exits of pills, unifying the counts for $W = 3, 4, 5, 6$. Here, the pair <stock ID, the time> is counted uniquely (one time for each pair for all $W$). The digits show the counts of cases where the stock prices increased/decreased from the period (of length $\Delta t$) before to the after each entrance/exit $t$. $\sigma$ means the standard deviation of the price of each stock for the period of length $\Delta t$ before $t$. The counted number may be smaller for a larger $\Delta t$ because the data period was the same ( Jul 2007 till Jan 2019).

| $\Delta t$ | Entrances (red) | | | | Exits (green) | | | |
|---|---|---|---|---|---|---|---|---|
| | 3 mo. | 6 mo. | 12 mo. | 24 mo. | 3 mo. | 6 mo. | 12 mo. | 24 mo. |
| decrease | 4(.11) | 7(.19) | 5(.14) | 4(.12) | 6(0.15) | 4(0.11) | 6(0.18) | 3(0.09) |
| increase | 32(.89) | 29(.81) | 29(.85) | 27(.88) | 34(.85) | 34(.89) | 28(.82) | 30(0.91) |
| increase > $\sigma$ | 27(.75) | 26(.72) | 27(.79) | 22(.71) | 24(.60) | 26(.69) | 23(.68) | 22(.67) |

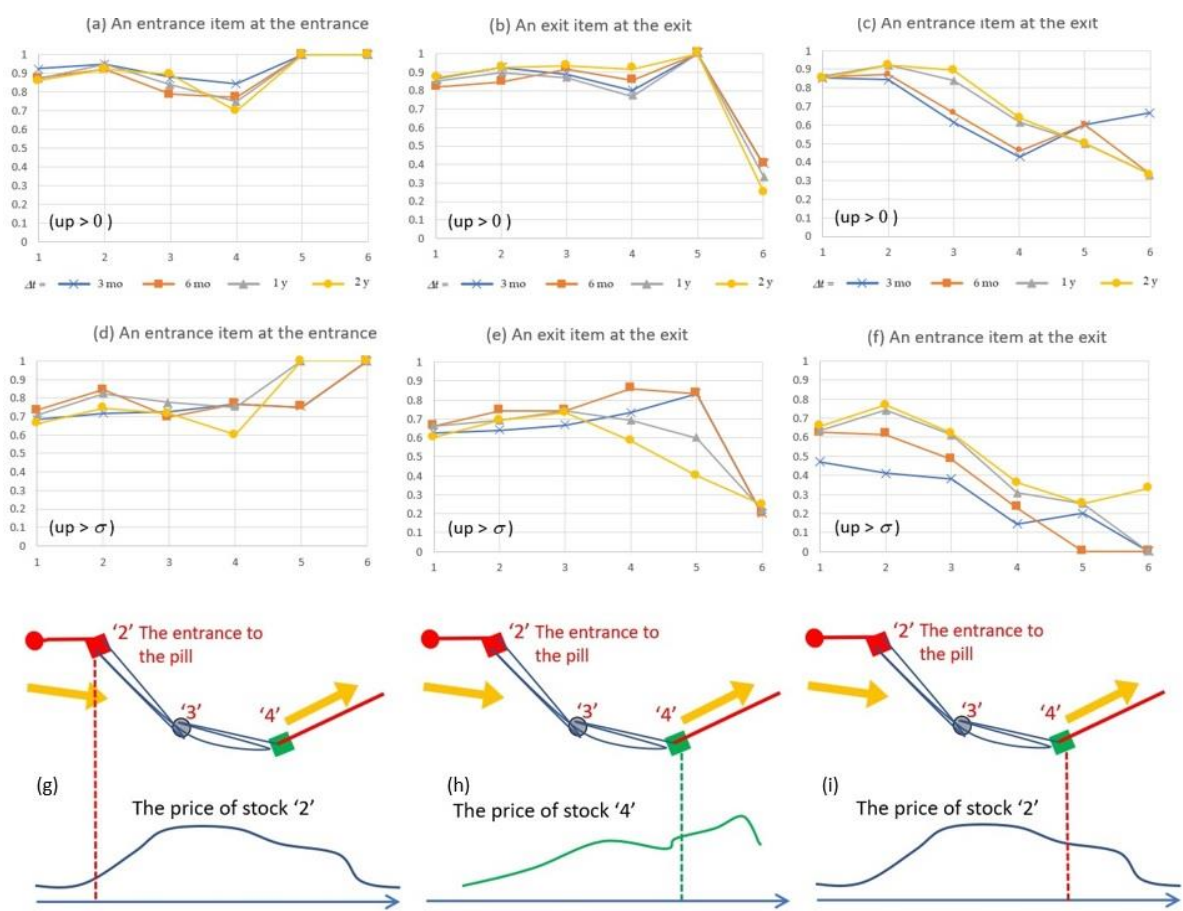

**Fig. 7** The price-up rates for entrance/exit items at the entrance/exits.

## 4. Discussions

Let us discuss the meaning of the results above from the viewpoint to fulfill the most essential purpose of this paper. That is, our aim is to aid the explanation of analysts in various time-scales about the market trend. In Figure 3 and 4, we find a high impact change from close to December 2012 (stock no.4404 in Figure 3 and no.4651 in Figure 4), corresponding to the political decision of Japan's Prime Minister Abe who started to promote the economy for the upward trend, called Abenomics. After this period, the upward trend has been reinforced by investors preferring to buy stocks of stable industry such as healthcare (3660, 4350), and energy (4651), IT (4229) that have been known to be low risk i.e., low volatility [26], because the governmental bond lost trust due to the lowered interest rate introduced by Bank of Japan since April 2013. Although the trend of software for smartphones diverting technologies developed for games and urban development based on big data came to be trendy according to the black letters in Figure 3, 4, and 5 showing the key events in pills (defined in subsection 2-A), experts (analysts and investors who listen to analysists) lead the market with their preference of low-risk stocks, that cause new trends. Here the larger letters (red: entrances of pills, green: exits) show events of the larger wire-weight defined in the algorithm. The low-volatility stocks did not cause new long-term trends to be shown as key events in pills, but contributed to the enhancement of innovative industries by triggering the opening of pills.

More about Figure 5, the largest (of the largest wire weight) green node 7022 is standing in July 2016 rather than the end of 2012. In the real events, the long-term interest rate came to be increased in Japan [27] from July 2016 and in US [28] from August 2016 in the context of low interest rate since April 2013 above. From the aspect of stock analysts, this is informative in explaining the scenario of general movements in the interest of investors. That is, after the beginning of Abenomics since 2012, due to the low rate of interest introduced by the government of Japan (minus rate at the beginning of 2016), investors who had been investing to governmental bonds shifted to buying low-volatility stocks such as specific electric/precision machineries, healthcare, and medical systems. These low-volatility stocks priced up since the end of 2012 and improved further in 2014 and 2016 according to stock analysts' comments about their general impression. Stocks of construction firms were also bought from a similar interest. On the other hand, because of the increase in the long-term interest rate in Japan and US since the July of 2016, as an effect opposite to the price-up of low-risk stocks as discussed before for the reduced interest rate due to Abenomics, high-risk stocks came to be bought relative to low-risk (i.e., low-volatility) stocks. As a result, the stocks on new industries such as digital devices for ultra-highspeed telecommunication, software for smart-phone with integrated use of big data (on maps and healthcare) and AI, and urban engineering for cities embracing innovative firms such as Shenzhen came to rise recently, as stock 6875 that is one of the few entrances of long ($W$=6) time-scale pills in Figure 6.

The results in Table 2 can be interpreted for a stock appearing at a pill entrance because the entrance means the beginning of the trend the stock is bought frequently. On the other hand, the exit stock may be intuitively felt to down price. However, the stock at an exit means to be of high impact enough to break the trend so far, so this tends to be a new focus of attention of investors. Rather, the decrease tendency is found at the exit of a pill for the stock that appeared as the entrance of the same pill. Other tendencies found in Figure 7, as mentioned in Section 3, are interpreted as follows.

(1) A trend starts at the entrance of a pill and continues during the pill. As a result, the price of the stock at the start (entrance of a pill) continues to increase during the trend (Figure 7(a), (d)).

(2) A trend ends at the exit of a pill, which may mean the impact of the stock at the ending point, but fades in a short time because the trend disappears with the pill disappears. As a result, the price of the stock at the ending point increases once but decreases sooner than stock at the entrance (Figure 7(b), (e)).

(3) A trend ends at the exit of a pill. As a result, the price of the stock at the entrance of a pill decreases sooner after the exit of the same pill than after the entrance (Figure 7(c), (f)).

(4) The stock tends to price up the more for the shorter $W$, because a short $W$ means a frequent repetition of the stock. However, even for a long $W$, the price appearing as the entrance increases continuously during the period of the trend started by itself (all in Figure 7).

Thus, we propose investors to consider to position all the stocks at the time they appear as entrances or as exits and sell the entrance stock by the time of the pill's exit or within the delay of length that is in a negative correlation to the length of the window width $W$.

## 5. Conclusions

Here we extended and applied TS to the data on the increase in stock prices, where patterns of various timescales co-exist and the boundaries of trends are hard to identify. Behind the sequence of stock prices, various investors really have different reasons for buying and/or selling stocks. By use of TS, the connections from/to trends in the market came to be visualized, to aid in explaining the changes in the market extracting patterns in the desired time scale. As a result, the changing in the prices of stocks came to be explained as a mixture of various time-scales, including middle-term changes caused by political decisions and long-term ones due to innovations in the industry. Also, it is found that the change points found as entrances to pills by TS coincide by higher precision with the changes in each stock prices, by as high accuracy of price increase of over 70%. This precision also varies by the timescale and the positioning (e.g., at the entrances/exits of pills or elsewhere) of the stocks in the string presented by TS, and the timing after the entrance to a pill. For the time being, the success of selling a stock at the exit of a pill is not as certain as buying the same stock when it started a pill, according to the probabilities of down pricing we obtained so far. In the future work, we are combining external knowledge or data to refine this certainty, as expected in the method $M_2$ in IMDJ.

The strength of TS is also in that the trend shifts in both of (1) the overall stock market and (2) the price of each stock can be explained with one string, whereas (1) and (2) have been analyzed or discussed separately in the previous work and the literature (e.g., [24] and [25] above). The direction to enable the analysis about the interrelationship between the global dynamics of the market and the local change in each stock is consistent with the fact that the former is composed of the latter and the latter is affected by the former. Thus, the ability to predict both with one output of visualization fits analysts' requirement for efficiency. As TS has been created from the requirement shown in IMDJ, this paper presents evidence that the product of the data market is of practical utility.

**Supplementary Materials:**

**Data Availability:** We used data on Japan's stock-price trend chart (Nikkei Average: https://indexes.nikkei.co.jp/en) for evaluating the performance of TS in detecting the up/down-ward price changes in the stock market. Such a scientific use of the data is allowed in https://indexes.nikkei.co.jp/nkave/archives/file/license_agreement_jp.pdf .

**Author Contributions:** conceptualization of Tangled String on IMDJ, Yukio Ohsawa and Teruaki Hayashi, methodology, software, Yukio Ohsawa;   validation,   Yukio Ohsawa Ohsawa and Takaaki Yoshino.

**Funding:** This work was funded by JST CREST No. JPMJCR1304, JSPS KAKENHI JP16H01836, and JP16K12428.

**Acknowledgments:** We appreciate for the stock analysts who provided comments on the results and the method of Tangled String adopted to stock analysis.

**Conflicts of Interest:** The authors declare no conflict of interest.